\nonstopmode

\newcommand{\ie}{i.e.{}}
\newcommand{\etal}{\textit{et al.}}
\newcommand{\U}[1]{\,{\rm{#1}}}
\newcommand{\I}[1]{_{\textrm{\scriptsize #1}}}
\newcommand{\Sum}{\sum\limits}
\newcommand{\angstrom}{\U{\hbox{\AA}}}
\newcommand{\degree}{{}^{\circ}}
\newcommand{\oHF}[1]{(HF)$_{#1}$}
\newcommand{\oHCl}[1]{(HCl)$_{#1}$}
\newcommand{\HF}{(HF)$_{\infty}$}
\newcommand{\HCl}{(HCl)$_{\infty}$}
\newcommand{\DCl}{(DCl)$_{\infty}$}
\newcommand{\Cal}[1]{{\cal #1}}
\newcommand{\mHartree}{\U{mE_{\mathit h}}}
\newcommand{\muHartree}{\U{\mu E_{\mathit h}}}
\newcommand{\bra}[1]{\left<\right.\!#1\!\left.\right|}
\newcommand{\ket}[1]{\left|\right.\!#1\!\left.\right>}
\newcommand{\euler}{\textrm{e}}

\documentclass[prb,showpacs,nopreprintnumbers,twocolumn]{revtex4}
\usepackage{graphicx,dcolumn,subeqnarray}

\begin{document}
\title{Hydrogen bonding in infinite hydrogen fluoride
and hydrogen chloride chains}
\date{October 14, 2006}
\author{Christian Buth}
\email[Corresponding author. Electronic address: ]{Christian.Buth@web.de}
\author{Beate Paulus}
\affiliation{Max-Planck-Institut f\"ur Physik komplexer Systeme, N\"othnitzer
Stra\ss{}e~38, 01187~Dresden, Germany}

\begin{abstract}
Hydrogen bonding in infinite~\HF{} and \HCl{} bent (zigzag) chains
is studied using the \emph{ab initio}
coupled-cluster singles and doubles~(CCSD) correlation method.
The correlation contribution to the binding energy is
decomposed in terms of nonadditive many-body interactions between
the monomers in the chains, the so-called energy increments.
Van der Waals constants for the two-body dispersion interaction
between distant monomers in the infinite chains
are extracted from this decomposition.
They allow a partitioning of the correlation contribution to
the binding energy into short- and long-range terms.
This finding affords a significant reduction in the computational
effort of \emph{ab initio} calculations for solids as only
the short-range part requires a sophisticated treatment
whereas the long-range part can be summed immediately
to infinite distances.
\end{abstract}

%
% The 2006 PACS classification scheme
%
% 31.15.Ar Ab initio calculations
% 31.25.Qm Electron correlation calculations for polyatomic molecules
% 71.15.Nc Total energy and cohesive energy calculations
% 71.20.Ps Other inorganic compounds
%

\preprint{arXiv:cond-mat/0601470}
\maketitle

\section{Introduction}

Hydrogen fluoride~\cite{Atoji:CS-54,Habuda:NM-71,Johnson:CS-75,%
Otto:BE-86,Panas:HF-93,Berski:DHF-98}
and hydrogen chloride~\cite{Sandor:CS-67,Sandor:CC-67,Sandor:ND-69}
are representatives of molecular crystals; the electronic structure
of the constituent HF or HCl~monomers is
essentially preserved upon crystallization.
The monomers in both crystals are hydrogen bonded;~\cite{Hamilton:HB-68,%
Pauling:NC-93,Scheiner:HB-97,Hadzi:TT-97,Karpfen:CE-02}
it is a directional and anisotropic bonding
of the hydrogen in a HF and HCl~monomer to the fluorine
or chlorine atom, respectively, of a neighboring monomer.
The bonding is caused by a partial withdrawal of charge from
the hydrogen atom due to the high electronegativity of
the fluorine or chlorine atoms.~\cite{Hamilton:HB-68,%
Pauling:NC-93,Scheiner:HB-97,Hadzi:TT-97,Karpfen:CE-02}
Hydrogen bonds are intermediates between ionic
bonding and van der Waals bonding and are of great
importance for the physical and chemical properties of many organic and
inorganic crystals. Moreover, they turn out to be crucial for the structure of
many biopolymers such as proteins and nucleic acids.~\cite{Hamilton:HB-68,%
Pauling:NC-93,Scheiner:HB-97,Hadzi:TT-97,Karpfen:CE-02}

\begin{figure}
  \includegraphics[width=\hsize,clip]{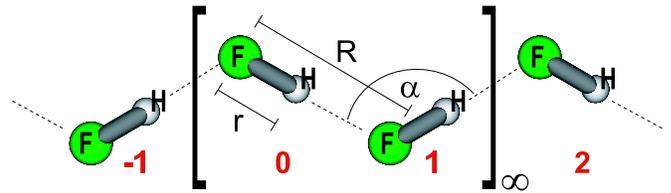}
  \caption{(Color online) Structure of infinite~\HF{} zigzag chains.
           The structure of \HCl{}~chains is analogous.
           The grey (red) numbers below the monomers indicate the relative
           position of a monomer in the chain with respect to the
           monomer~``0'' at the origin.}
  \label{fig:polygeom}
\end{figure}

At low temperature, HF and HCl crystals are structurally
very similar.
While HF forms strong hydrogen bonds, HCl forms
weak hydrogen bonds. Therefore, HF and HCl
represent good candidates for
a thorough analysis of this special type of bonding in crystals.
The monomers in both compounds are found to be arranged in terms of
parallel zigzag chains with a large interchain distance and, hence,
a weak interchain interaction [see Sec.~\ref{sec:comp} for details].
Frequently, a single infinite chain is considered as a simple but
realistic model of the crystals.
The isolated \HF{}~chain has fascinated many theoreticians
and much work has been carried out dealing with it;
namely, model studies,~\cite{Springborg:ES-87,Springborg:ES-88}
semiempirical~(intermediate neglect of differential overlap)
examinations of~\HF{} by Zunger,~\cite{Zunger:BS-75}
density functional theory calculations~[local density
approximation] of Springborg,~\cite{Springborg:ES-87,Springborg:ES-88}
and \emph{ab initio}
investigations.~\cite{Kertesz:AI-75,Karpfen:AI-76,%
Blumen:CC-76,Blumen:EB-77,Kertesz:AI-78,Karpfen:AI-80,Karpfen:HB-81,%
Karpfen:AI-82,Ihaya:AI-84,Liegener:AI-87,%
Mayer:CH-97,Berski:PHF-97,Hirata:AI-98,Jacquemin:LR-99,Buth:BS-04,Buth:MC-05}
In contrast, the \HCl{} chain has not been investigated very extensively;
there are only a few \emph{ab initio} examinations.~\cite{Blumen:CC-76,Blumen:EB-77,Blumen:EBS-77,Berski:PHF-98}
For both isolated chains the zigzag geometry of Fig.~\ref{fig:polygeom}
is the energetically favored arrangement in comparison with
the linear geometry. For \HF{} chains this has been shown in
early studies of Karpfen~\etal~\cite{Karpfen:AI-76,Karpfen:AI-80} and
Beyer and Karpfen.~\cite{Karpfen:AI-82}
The structure of isolated~\HCl{} chains is investigated
in Ref.~\onlinecite{Berski:PHF-98}.

The large number of \emph{ab initio} studies that have
been carried out for the chains also exhibits how
challenging hydrogen-bonded systems are.
An accurate treatment of hydrogen bonds requires both a
good electron correlation method and a large one-particle
basis set.~\cite{Halkier:BS-99,Buth:BS-04,Buth:MC-05}
Hence, in Refs.~\onlinecite{Buth:BS-04,Buth:MC-05}, we applied
basis set extrapolation schemes---which
are powerful methods to improve the accuracy
of both Hartree-Fock and correlation
energies for small molecules---to extended systems,
showing their validity and usefulness also in
this situation.
Thereby, a periodic Hartree-Fock treatment
was accompanied by correlation calculations.

We use the coupled-cluster singles and doubles~(CCSD) correlation
method~\cite{Purvis:CC-82} in conjunction with the
incremental scheme~\cite{Stoll:CD-92,%
Stoll:CS-92,Stoll:CG-92,Fulde:EC-95,Fulde:WF-02}
which has been used in many applications and yields
a physically meaningful many-body decomposition
of the correlation energy.
It has been applied to a few one-dimensional systems before: the
\textit{trans}-polyacetylene polymer,~\cite{Yu:IA-97}
the infinite lithium hydride chain,~\cite{Abdurahman:AI-00}
the beryllium hydride polymer,~\cite{Abdurahman:AI-00} and
poly(\textit{para}-phenylene).~\cite{Willnauer:QC-04}
In the context of the present paper, the contributions of
Doll~\etal,~\cite{Doll:CE-95,Doll:CE-96,Doll:CP-97,Doll:GS-98}
who studied ionic solids, as well as the investigations of
rare-gas crystals by Ro\'sciszewski~\etal~\cite{Rosciszewski:AI-99,%
Rosciszewski:AI-00} are worth mentioning because
hydrogen-bonded crystals fall in between these two
types of bonding.

This paper is structured as follows.
The CCSD electron correlation method and the incremental scheme
are introduced in Sec.~\ref{sec:theory} whereas Sec.~\ref{sec:comp}
describes geometries, basis sets, and the computer programs employed.
In Sec.~\ref{sec:elcorr}, we discuss electron
correlation effects and their implications.
The contributions to the binding energy of
the~\HF{} and \HCl{} chains are evaluated in
Sec.~\ref{sec:bind} and conclusions are drawn in
Sec.~\ref{sec:conclusion}.

\section{Theory}
\label{sec:theory}

We choose the Hartree-Fock approximation as a starting point to
study electron correlations, employing the full non-relativistic
Hamiltonian of the chains within the fixed-nuclei
approximation.~\cite{Szabo:MQC-89,Fulde:EC-95,Ladik:PS-99,%
Fulde:WF-02,Buth:MC-05}
The resulting Hartree-Fock Bloch orbitals are transformed
to Wannier orbitals because the latter provide a more
appropriate, local representation of the Hamiltonian for
a subsequent treatment of electron
correlations.~\cite{Fulde:EC-95,Fulde:WF-02,Forner:LC-92}
The ground-state wave function is modeled by the coupled-cluster
ansatz~\cite{Szabo:MQC-89}
\begin{equation}
  \ket{\Psi_0^N} = \euler^{\hat T} \ket{\Phi_0^N}
\end{equation}
which relates the $N$-electron Hartree-Fock ground-state wave
function~$\Phi_0^N$ to the correlated ground-state wave
function~$\Psi_0^N$ employing the cluster operator~$\hat T$.
For the calculations in this study,
it is sufficient to restrict $\hat T$~to single and double
excitations, resulting in the coupled-cluster singles and
doubles~(CCSD) scheme~\cite{Purvis:CC-82,Hirata:CC-03}
\begin{eqnarray}
  \hat T &=& \Sum_{\vec R_1^{\prime} \, \alpha_1 \atop \vec R_1 \, \kappa_1}
    t^{\vec R_1^{\prime} \, \alpha_1}_{\vec R_1 \, \kappa_1}
    \hat c^{\dagger}_{\vec R_1^{\prime} \, \alpha_1}
    \hat c_{\vec R_1 \, \kappa_1} \\
  &&{} + \Sum_{\vec R_1^{\prime} \, \alpha_1 < \vec R_2^{\prime} \, \alpha_2
    \atop \vec R_1 \, \kappa_1 < \vec R_2 \, \kappa_2}
    t^{\vec R_1^{\prime} \, \alpha_1, \vec R_2^{\prime} \, \alpha_2}_{\vec
       R_1 \, \kappa_1, \vec R_2 \, \kappa_2}
    \hat c^{\dagger}_{\vec R_1^{\prime} \, \alpha_1}
    \hat c^{\dagger}_{\vec R_2^{\prime} \, \alpha_2} \hat c_{\vec R_1 \, \kappa_1}
    \hat c_{\vec R_2 \, \kappa_2}
    \; . \nonumber
\end{eqnarray}
The coefficients~$t^{\vec R_1^{\prime} \, \alpha_1}_{\vec R_1 \, \kappa_1}$
and $t^{\vec R_1^{\prime} \, \alpha_1, \vec R_2^{\prime} \, \alpha_2}_{\vec
R_1 \, \kappa_1, \vec R_2 \, \kappa_2}$~are referred to as excitation
amplitudes.
Let~$\vec r$ and $s$ denote spatial and spin coordinates;
then $\hat c^{\dagger}_{\vec R_1^{\prime} \, \alpha_1}$ creates electrons
in virtual spin Wannier orbitals~$w_{\vec R_1^{\prime} \, \alpha_1}(\vec
r \, s)$ whereas $\hat c_{\vec R_1 \, \kappa_1}$ annihilates
electrons from occupied spin Wannier orbitals~$w_{\vec R_1 \,
\kappa_1}(\vec r \, s)$.
Here $\vec R_1$ and $\vec R_1^{\prime}$ denote the unit cell in
which the Wannier orbital is located and $\alpha_1$ and
$\kappa_1$ refer to Wannier orbital indices.
Assuming Born-von K\'arm\'an boundary conditions,
the correlation energy per unit
cell~$\Cal E_{\rm corr}$ is given by
\begin{equation}
  \label{eq:ccsdenergy}
  N_0 \, \Cal E_{\rm corr} = \bra{\Psi_0^N} \hat H \ket{\Psi_0^N}
    - \bra{\Phi_0^N} \hat H \ket{\Phi_0^N} \; ,
\end{equation}
where $N_0$~is the number of unit cells in the chains.

Occupied Wannier orbitals are grouped in terms of $n_{\rm one}$
pairwise disjunct one-body orbital sets which are defined
by
\begin{equation}
  \label{eq:onebset}
  \vec R \, I_l = \{w_{\vec R \, \alpha}(\vec r \, s) \> |
    \> \alpha \in I_l\}
\end{equation}
for~$l = 1, \ldots, n_{\rm one}$.

The expression for the correlation energy of a
crystal~(\ref{eq:ccsdenergy}) is rearranged and decomposed in terms
of correlation energies~$\varepsilon_{\vec R_1 \, I_1 \cdots
\vec R_K \, I_K}$ of the electrons from
the one-body orbital sets~$\vec R_1 \, I_1 \cdots \vec R_K \, I_K$.
The resulting formula for the correlation energy of the chains
per unit cell~$\Cal E_{\rm corr}$ reads~\cite{Stoll:CD-92,%
Stoll:CS-92,Stoll:CG-92,Fulde:EC-95,Yu:IA-97,Abdurahman:AI-00,%
Fulde:WF-02,Willnauer:QC-04,Buth:MC-05}
\begin{widetext}
\begin{equation}
  \label{eq:inccryst}
  \begin{array}{rcl}
    N_0 \, \Cal E_{\rm corr} &=& \frac{1}{1!} \, \Sum_{\vec R_1 \, I_1}
      \Delta \varepsilon_{\vec R_1 \, I_1} + \frac{1}{2!} \, \Sum_{\vec R_1 \,
      I_1 \neq \vec R_2 \, I_2} \Delta \varepsilon_{\vec R_1 \, I_1 \; \vec R_2
      \, I_2} \\
    &&{} + \cdots + \frac{1}{K!} \, \Sum_{\vec R_1 \, I_1 \cdots \vec R_K \, I_K
      \atop \textrm{\scriptsize pairwise disjunct}}
      \Delta \varepsilon_{\vec R_1 \, I_1 \cdots \vec R_K \, I_K}
      + \cdots \; . \nonumber \\
\end{array}
\end{equation}
The one-body~$\Delta \varepsilon_{\vec R_1 \, I_1}$, the
two-body~$\Delta \varepsilon_{\vec R_1 \, I_1 \; \vec R_2 \, I_2}$,
\dots, up to the $K$-body~$\Delta \varepsilon_{\vec R_1 \, I_1 \cdots
\vec R_K \, I_K}$,~\ldots{} energy increments are defined recursively
by~\cite{Stoll:CD-92,Stoll:CS-92,Stoll:CG-92,Fulde:EC-95,Yu:IA-97,%
Abdurahman:AI-00,Fulde:WF-02,Willnauer:QC-04,Buth:MC-05}
\begin{subeqnarray}
  \label{eq:incdefs}
  \Delta \varepsilon_{\vec R_1 \, I_1} &=& \varepsilon_{\vec R_1 \, I_1} \; , \\
  \Delta \varepsilon_{\vec R_1 \, I_1 \; \vec R_2 \, I_2} &=&
  \varepsilon_{\vec R_1 \, I_1 \; \vec R_2 \, I_2} - \Delta
    \varepsilon_{\vec R_1 \, I_1} - \Delta \varepsilon_{\vec R_2 \, I_2} \; , \\
  & \vdots & \nonumber \\
  \Delta \varepsilon_{\vec R_1 \, I_1 \cdots \vec R_K \, I_K} &=&
    \varepsilon_{\vec R_1 \, I_1 \cdots \vec R_K \, I_K}
    - \Sum_{n = 1}^{K - 1}
    \frac{1}{n!} \Sum_{{{
    \{ \vec R_1^{\prime} \, I_1^{\prime} \cdots \vec R_n^{\prime}
    \, I_n^{\prime} \}
    \atop \subset \{ \vec R_1 \, I_1 \cdots \vec R_K \, I_K \} }
    \atop \vec R_1^{\prime} \, I_1^{\prime} \cdots \vec R_n^{\prime}
    \, I_n^{\prime} }
    \atop \textrm{\scriptsize pairwise disjunct}}
    \Delta \varepsilon_{\vec R_1^{\prime} \, I_1^{\prime} \cdots \vec
    R_n^{\prime} \, I_n^{\prime}} \; . \\
  & \vdots & \nonumber
\end{subeqnarray}
\end{widetext}
The factors~$\frac{1}{1!}$, $\frac{1}{2!}$, \ldots, $\frac{1}{K!}$,
\ldots{} in front of the sums on the right-hand side of
Eq.~(\ref{eq:inccryst}) account for permutations among the one-body
orbital sets of a certain $K$-body energy increment~($K \geq 2$).
One can eliminate all permutations that lead to the
same energy increment by letting the sums in Eq.~(\ref{eq:inccryst})
run only over distinct sets of one-body orbital sets.

The translational relation of the Wannier orbitals
leads to the translational symmetry of the energy
increments, \ie, $\Delta \varepsilon_{\vec 0 \, I_1 \cdots \vec R_K
- \vec R_1 \, I_K} = \Delta \varepsilon_{\vec R_1 \, I_1 \cdots \vec R_K
\, I_K}$. This can be exploited in Eq.~(\ref{eq:inccryst})
to make the right-hand side of Eq.~(\ref{eq:inccryst})
independent of the first lattice sum~$\Sum_{\vec R_1}$ which, for this
reason, is $N_0$~times the sum of the translational
symmetry-adapted energy increments.
It allows us to eliminate the factor~$N_0$ in front of the
left-hand side of Eq.~(\ref{eq:inccryst})
and enormously reduces the number of energy increments
that need to be calculated to describe the chains with a
given accuracy.

The localized occupied molecular orbitals of oligomers, comprising a few
monomers arranged in the geometry of the infinite chains, are
found to be a good approximation to the Wannier orbitals of infinite
chains and crystals.~\cite{Stoll:CD-92,Stoll:CS-92,Stoll:CG-92,%
Fulde:EC-95,Yu:IA-97,Abdurahman:AI-00,Fulde:WF-02,Willnauer:QC-04}
This facilitates the determination of the correlation contribution
to the binding energy of infinite chains from
the energy increments in oligomers using Eq.~(\ref{eq:inccryst}).
Yet the virtual Wannier orbitals involved in the determination
of the energy increments, and implicitly contained in
Eq.~(\ref{eq:incdefs}), are replaced by all the virtual canonical
molecular orbitals of the oligomers.
The procedure outlined in this paragraph is termed the incremental
scheme.~\cite{Stoll:CD-92,Stoll:CS-92,Stoll:CG-92,%
Fulde:EC-95,Fulde:WF-02}

\section{Computational details}
\label{sec:comp}

Hydrogen fluoride and hydrogen chloride crystallize in an orthorhombic
low-temperature phase described by the space groups~$Bm2_1b$ for
HF,~\cite{Atoji:CS-54,Habuda:NM-71,Johnson:CS-75} and
$Bb2_1m$ for HCl.~\cite{Sandor:CS-67,Sandor:CC-67,Sandor:ND-69}
The unit cells of both crystals contain four monomers which
are arranged in terms of weakly interacting parallel zigzag
chains~[Fig.~\ref{fig:polygeom}].
The chains are described by a unit cell which comprises two monomers
and are considered as an excellent one-dimensional model for
HF and HCl crystals.~\cite{Zunger:BS-75,%
Kertesz:AI-75,Karpfen:AI-76,Blumen:CC-76,%
Blumen:EB-77,Blumen:EBS-77,Kertesz:AI-78,%
Karpfen:AI-80,Karpfen:HB-81,Karpfen:AI-82,%
Ihaya:AI-84,Liegener:AI-87,%
Springborg:ES-87,Springborg:ES-88,Mayer:CH-97,%
Berski:PHF-97,Berski:PHF-98,Hirata:AI-98,%
Jacquemin:LR-99,Buth:BS-04,Buth:MC-05}
The structure of a single chain is determined by
three parameters, the H---$X$~distance~$r$,
the $X \cdots X$~distance~$R$, and the
angle~$\alpha = \angle({\rm H}X{\rm H})$, $X ={}$F$,$Cl.
Experimental values for the parameters are~$r = 0.92 \angstrom$,
$R = 2.50 \angstrom$, and $\alpha = 120 \degree$ for
\HF{},~\cite{Atoji:CS-54,Holleman:IC-01} and
$r = 1.25 \angstrom$, $R = 3.688 \angstrom$, and
$\alpha = 93.3 \degree$ for
\DCl{}.~\cite{Sandor:CS-67} HCl and deuterated DCl crystals
have very similar lattice
constants and are considered to be isomorphous.~\cite{Sandor:CS-67}
Unfortunately, further structural information for HCl crystals
is unavailable.

To calculate the energy increments in Eq.~(\ref{eq:inccryst}), we
perform molecular calculations on oligomers~\oHF{n} and
\oHCl{n}, respectively, \ie, short fragments of the chains.
We utilize the program package \textsc{molpro},~\cite{MOLPRO2002.6}
employing Foster-Boys localization~\cite{Boys:MO-60,Foster:MO-60}
of the molecular orbitals and the
coupled-cluster singles and doubles~(CCSD) correlation
method.~\cite{Purvis:CC-82,Hampel:CE-92,MOLPRO2002.6}
Thereby, we are consistent with Refs.~\onlinecite{Buth:BS-04,Buth:MC-05}
where also the CCSD~method was employed.
We use the aug-cc-pVDZ basis set~\cite{Dunning:GBS-89,%
Kendall:EA-92,Woon:GBS-93,basislib-04} which
yields somewhat less accurate total
binding energies than those obtained in
Refs.~\onlinecite{Buth:BS-04,Buth:MC-05}.
Yet we are mainly interested in the long-range behavior of
electron correlations in the hydrogen-bonded chains;
it is well described by the aug-cc-pVDZ basis set due
to vanishingly small geometrical orbital
overlaps among the one-body orbital sets
in the $K$-body energy increments considered.
In fact, the modulus of the two-body energy increment
involving the third-nearest-neighbor monomer~$|\varepsilon_{03}|$,
as obtained with the aug-cc-pVDZ basis set,~\cite{Dunning:GBS-89,%
Kendall:EA-92,Woon:GBS-93,basislib-04} is smaller
%
% HF: EINC47 aug-cc-pVDZ - aug-cc-pVTZ
% (-0.00000737 - (-0.00000752)) / 0.00000737 = 2.0%
%
by~$2 \%$ for~\HF{} and
%
% HCl: EINC36 aug-cc-pVDZ - aug-cc-pVTZ
% (-0.00001059 - (-0.00001106)) / -0.00001059 = 4.4%
%
by~$4 \%$ for~\HCl{} than~$|\varepsilon_{03}|$
obtained with the larger aug-cc-pVTZ basis set~\cite{Dunning:GBS-89,%
Kendall:EA-92,Woon:GBS-93,basislib-04}
[Sec.~\ref{sec:elcorr}].
Similarly, $K$-body energy increments for $K \geq 3$ agree
well as soon as the geometrical orbital overlaps between the
one-body orbital sets become negligible.

In Tab.~\ref{tab:bindtotal} below, reference is made to periodic
Hartree-Fock calculations of~\HF{} and \HCl{}
chains, described in Refs.~\onlinecite{Buth:BS-04,Buth:MC-05},
which were carried out with the \textsc{crystal}
program.~\cite{Pisani:HF-88,Pisani:QM-96,crystal03}
For these calculations, the $f, g, h, i$ functions
were removed from the basis sets.
To estimate the influence of the neglected basis functions on the
total binding energy of the chains, we use
\textsc{molpro},~\cite{MOLPRO2002.6} to calculate the Hartree-Fock
binding energy per monomer for~\oHF{9} and \oHCl{9}
both with and without $f, g, h$~functions using the
cc-pV5Z basis set;~\cite{Dunning:GBS-89,Woon:GBS-93,basislib-04}
at the Hartree-Fock level it is close to completeness with
respect to basis functions with the angular momenta~$s, p, d$.
The binding energies of both chains
obtained with $f, g, h$~functions are smaller
%
% Estimate Hartree-Fock binding in (HF)_9 energy of cc-pV5Z minus
%       cc-pV5Z-without-f,g,h-functions
%
% HF
% cc-pV5Z        : -900.70115997 / 9 - (-100.07035200) = -0.00755466
% cc-pV5Z-crystal: -900.69710425 / 9 - (-100.06988760) = -0.00756843
% cc-pV5Z minus cc-pV5Z-crystal = -0.00755466 - (-0.00756843) = 0.00001377
%
%
% HCl
% cc-pV5Z        : -4141.01380657 / 9 - (-460.11238499) = -0.0002602
% cc-pV5Z-crystal: -4141.00858712 / 9 - (-460.11179208) = -0.0002732
% cc-pV5Z minus cc-pV5Z-crystal = -0.00026020 - (-0.00027320) = 0.00001300
%
by~$\approx 13 \U{\mu E_h}$ than those determined without them.
This deviation turns out to be sufficiently small to be neglected
in the subsequent calculation of total binding energies
in Sec.~\ref{sec:bind}.
However, note that for the correlation calculations
the $f, g, h$~functions are fully considered.

The calculated binding energies reported here
account for the basis set superposition
error~\cite{Duijneveldt:CP-94}~(BSSE)
by the counterpoise correction~(CP);
it is determined by surrounding an isolated HF or HCl
monomer by sufficiently many ghost atoms positioned in
the geometry of the infinite
chains.~\cite{Boys:CP-70,Duijneveldt:CP-94}

\section{Electron correlations}
\label{sec:elcorr}
\subsection{Transferability}
\label{sec:transfer}

We have to ensure that the Foster-Boys-localized
molecular orbitals~\cite{Boys:MO-60,Foster:MO-60}
which are extracted from oligomers
are a good approximation to the Wannier orbitals of the
infinite \HF{} and \HCl{}~chains.
They have to be approximately translationally
related within a certain oligomer and
must not differ substantially
among oligomers of varying length.
These two properties of the localized
occupied molecular orbitals are termed transferability.
In order to achieve transferability for
moderately sized molecular clusters, one is
frequently obliged to account for the omitted
monomers in terms of an appropriate substitute.
In the case of~\HF{} and \HCl{}, two options
offer themselves. On the one hand, both
chains are cut out from a molecular crystal, implying
no substitutes as in rare-gas
crystals;~\cite{Rosciszewski:AI-99,Rosciszewski:AI-00}
on the other hand, particularly~\HF{} is rather
ionic and a surrounding by point charges
can be envisaged to create the embedding
which adequately models the infinite
chains.~\cite{Evjen:OS-32,Doll:CE-95,%
Doll:CE-96,Doll:CP-97,Doll:GS-98}
I explore both possibilities to elucidate what kind of procedure is
adequate for hydrogen-bonded solids.
The virtual orbitals of the oligomers are not altered
and remain canonical molecular orbitals
in the calculation of the individual energy increments.
The oligomer approximation thus corresponds
to a sort of domain decomposition of the
virtual space which also is applied in other
local correlation methods.~\cite{Pisani:LT-03}

\begin{figure}
  \includegraphics[width=\hsize,clip]{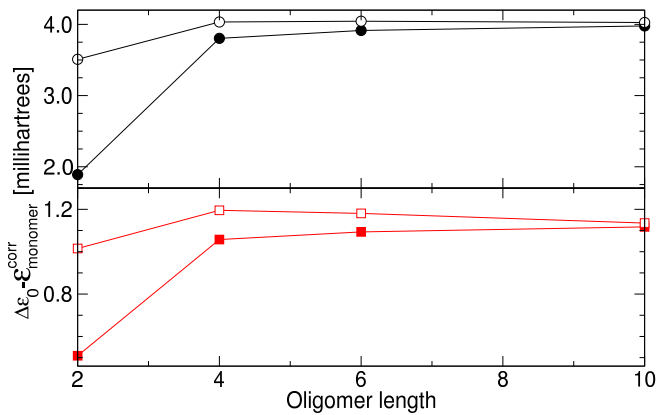}
   \caption{(Color online) One-body energy increments~$\Delta\varepsilon_0$
            of~\HF{} and \HCl{} chains as determined in oligomers of varying
            length reduced by the CP-corrected correlation energy~$\Cal
            E\I{mon}^{\rm corr}$ of the corresponding isolated monomer.
            The mean value of the two innermost energy increments
            of the oligomers is taken.
            Circles stand for~\HF{} and squares
            for~\HCl{} data. Open symbols refer to
            oligomers surrounded by point charges
            whereas closed symbols denote isolated
            oligomers.}
  \label{fig:one-body}
\end{figure}

\begin{figure}
  \includegraphics[width=\hsize,clip]{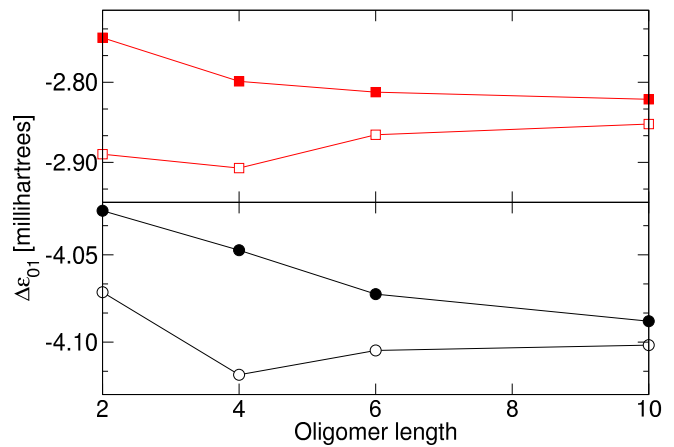}
   \caption{(Color online) Two-body energy increments~$\Delta\varepsilon_{01}$
            of~\HF{} and \HCl{} chains as determined in oligomers of varying length.
            The two innermost one-body orbital sets of the respective
            oligomer are taken to determine~$\Delta\varepsilon_{01}$.
            Symbols are chosen as in Fig.~\ref{fig:one-body}.}
  \label{fig:two-body}
\end{figure}

In the case of~\HF{} and \HCl{}, the outer and inner valence
orbitals, \ie, eight electrons, are assigned to the one-body
orbital set representing a particular HF or HCl~monomer~(\ref{eq:onebset}).
To test the transferability, we
calculate all one-body energy increments
and all two-body energy increments between
two adjacent monomers, the so-called
connected two-body energy increments,
in isolated oligomers~\oHF{n} and
\oHCl{n} for $n = 2, 4, 6, 10$.
They are compared with the energy increments from the
oligomers which were surrounded by point charges
up to twentieth-nearest neighbors.
The hydrogen atoms are represented by~$q = +1$;
fluorine and chlorine atoms are described
by~$q = -1$ with the exception of the terminal charges
which are set to~$q = -\frac{1}{2}$.~\cite{Evjen:OS-32,%
Doll:CE-95,Doll:CE-96,Doll:CP-97,Doll:GS-98}

Comparing the one-body and the connected two-body energy
increments obtained in oligomers of varying length in
Figs.~\ref{fig:one-body} and \ref{fig:two-body}, we
observe a rapid convergence of their values toward the
limit of the infinite chains
where the values determined for the isolated
oligomers and the values resulting from
the point-charge-embedded oligomers
approach each other quickly.
The energy increments taken from the isolated oligomers
increase (decrease) in Fig.~\ref{fig:one-body}~(Fig.~\ref{fig:two-body})
monotonically with the oligomer length
whereas the embedding with point charges causes an unsystematic
but quicker convergence.
The convergence behavior of the individual curves
in Figs.~\ref{fig:one-body} and \ref{fig:two-body}
is governed by a small variation in the localized orbitals
and, predominantly, by the improvement of the electronic
structure which approaches rapidly
the electronic structure of the infinite chains with
increasing length of the oligomers.
Obviously, the decision whether to employ point
charges or not has only a small impact on the values of the
energy increments.
The one-body (two-body) energy increments differ by
%
% CCSD, aug-cc-pVDZ, (HF)_10
% EINC4 - EINC4pointcharges
% -0.22338000 - (-0.22331345) = -0.00006655 Eh = -67 muEh
%
$67 \muHartree$
%
% CCSD, aug-cc-pVDZ, (HF)_10
% EINC45 - EINC45pointcharges
% -0.00408785 - (-0.00410198) =  0.00001413 Eh =  14 muEh
%
($14 \muHartree$) in~\HF{} and by
%
% CCSD, aug-cc-pVDZ, (HCl)_10
% EINC4 - EINC4pointcharges
% -0.175099 - (-0.17507709) = -0.00002191 Eh = -21.91 muEh
%
$22 \muHartree$
%
% CCSD, aug-cc-pVDZ, (HCl)_10
% EINC45 - EINC45pointcharges
% -0.00282125 - (-0.00285217) = 0.00003092 Eh = 30.92 muEh
%
($31 \muHartree$) in~\HCl{}.
In what follows only energy increments taken from
isolated oligomers are considered, as in
Refs.~\onlinecite{Buth:BS-04,Buth:MC-05}.

The error on the binding
energy due to the approximate translational
relation of the Wannier orbitals
taken from the finite-length oligomers can be estimated
by investigating the numerical
differences between energy increments whose values should
be identical due to (translational) symmetry. The
connected two-body increments, for example,
vary at most by~$40 \U{\mu E_h}$ in~\oHF{10} and at most
by~$11 \U{\mu E_h}$ in~\oHCl{10}.%
~\footnote{The two first and the two last monomers at the
ends of the oligomers are not regarded as they are
appreciably influenced by finite-size effects.}
We conclude that the transferability prerequisite
of the energy increments is satisfied within good accuracy.
The agreement between the results obtained in the two
different ways described before corroborates
the applicability of oligomers to represent the
electronic structure of infinite~\HF{}
and \HCl{} chains.

\begin{table}
  \centering
  \begin{ruledtabular}
    \begin{tabular}{ldd}
      Increment     & $\HF$ & $\HCl$ \\
      \hline
      \quad $\Delta\varepsilon_{0}$          & -223.3632 & -175.0985 \\
      \quad $\Delta\varepsilon_{0\,1}$       &   -4.0879 &   -2.8213 \\
      \quad $\Delta\varepsilon_{0\,2}$       &   -0.0861 &   -0.1347 \\
      \quad $\Delta\varepsilon_{0\,1\,2}$    &   -0.0163 &    0.0014 \\
      \quad $\Delta\varepsilon_{0\,3}$       &   -0.0063 &   -0.0092 \\
      \quad $\Delta\varepsilon_{0\,1\,3}$    &   -0.0015 &   -0.0016 \\
      \quad $\Delta\varepsilon_{0\,1\,2\,3}$ &    0.0003 &    0.0006 \\
    \end{tabular}
  \end{ruledtabular}
  \caption{Exemplary energy increments of~\HF{} and \HCl{} chains
           taken from~\oHF{10} and \oHCl{10} oligomers.
           The energy increments are computed using the one-body
           orbital sets of the innermost monomers in the oligomers.
           All data are given in millihartrees.}
  \label{tab:increments}
\end{table}

\subsection{Short-range correlations}
\label{sec:increments}

Let us discuss the short-range correlation contributions first.
The most relevant energy increments of~\HF{}
and \HCl{} are summarized in Tab.~\ref{tab:increments}.
The CP-corrected correlation energy of the
%
% aug-cc-pVDZ, CP in (HF)_9
% -100.261311923055 - (-100.033971301475) = -0.227340621580
%
HF monomer, $-227.3406 \mHartree$,
%
% aug-cc-pVDZ, CP in (HCl)_9
% -460.268686021020 - (-460.092470961000) = -0.176215060020
%
and of the HCl monomer, $-176.2151 \mHartree$, are
%
% aug-cc-pVDZ, HF
% (-0.227340621580 - (-0.223363216268)) / (-0.223363216268)
%       = 0.01780689487935986815
%
only larger in magnitude by~$1.8 \%$
%
% aug-cc-pVDZ, HCl
% (-0.176215060020 - (-0.175098524764)) / (-0.175098524764)
%       = -0.00637661
%
and $0.6 \%$, respectively, than the
corresponding one-body energy
increments~$\Delta\varepsilon_0$.
They are a bit larger because the electrons of a monomer
in the chains experience Pauli repulsion exerted by the
neighboring monomers.
However, these effects are small in our case and the electronic
structure of the~HF and HCl monomers is essentially
preserved in the \HF{} and \HCl{}~chains.

The $K$-body energy increments, $K \geq 2$, describe the mutual
correlation of the valence electrons of several monomers.
They give rise to a pronounced nonlinear increase of the binding energy
of small clusters and short oligomers which is termed bond
cooperativity (Ref.~\onlinecite{Karpfen:CE-02}
and references therein).
The modulus of the connected two-body energy increment
of~\HCl{}, $|\Delta\varepsilon_{01}^{\rm HCl}|$,
%
% aug-cc-pVDZ
% (-4.0879 - (-2.8213)) / (-4.0879) = 0.309841
%
is~$31 \%$ smaller than the modulus of the corresponding
energy increment of~\HF{}, $|\Delta\varepsilon_{01}^{\rm HF}|$.
The connected three-body energy
increment~$\Delta\varepsilon_{012}^{\rm HCl}$
is repulsive, \ie, greater than zero, but~$\Delta\varepsilon_{012}^{\rm HF}$
is again attractive, \ie, smaller than zero.
The reverse trend is observed for the remaining energy increments.
The modulus of the energy increment~$\Delta\varepsilon_{02}^{\rm HCl}$
%
% aug-cc-pVDZ
% (-0.0861 - (-0.1347)) / (-0.0861) = -0.56446
%
is~$57 \%$ larger than~$|\Delta\varepsilon_{02}^{\rm HF}|$.
The connected four-body increment~$\Delta\varepsilon_{0123}^{\rm HCl}$
%
% aug-cc-pVDZ
% (0.0006 - (0.0003)) / (0.0003) = 1.0
%
is~$100 \%$ larger than~$\Delta\varepsilon_{0123}^{\rm HF}$ and both
are repulsive.
These two trends of the energy increments can be explained
by two effects.
First, short-range correlation is effective for nearest
neighbors [see also the ensuing Sec.~\ref{sec:long}];
it is apparently stronger in~\HF{} than in \HCl{} due to
the tighter packing of the monomers in~\HF{}
and the higher compactness of the HF~monomer itself.
Second, chlorine atoms have a higher polarizability than
fluorine atoms because their valence electrons
are more diffuse than those of the latter atom.
This causes the van der Waals interaction
to be stronger in~\HCl{} than in~\HF{},~\cite{Craig:MQ-84}
leading to the more distant energy increments being larger in~\HCl{}
when compared with~\HF{};
all this despite the larger intermonomer distances
in~\HCl{} than in~\HF{}.

However, note that one should not associate too much physics with the
particular values of energy increments;
namely, they depend on the specific unitary transformation
used to localize the occupied orbitals
(here the one of Foster and Boys~\cite{Boys:MO-60,Foster:MO-60}).
Only the binding energy, which involves the sum of all the energy
increments [Eqs.~(\ref{eq:RHF_bind}) and (\ref{eq:corrpoly})],
is a physical observable and thus invariant under
orbital transformations.
When an analysis of the individual energy increments
is to be meaningful, their value should be
fairly independent of the localization procedure
which is well satisfied in our case.

\begin{table}
  \centering
  \begin{ruledtabular}
    \begin{tabular}{rddd}
      Compound & C_6^{\rm odd} & C_6^{\rm even} & $Experiment$%
\footnote{The experimental data are taken from Ref.~\onlinecite{Kumar:PS-85};
it is a rotational average of the van der Waals constants for the
dispersion interaction between two isolated monomers.} \\
      \hline
      % Cardinal number fit
      \HF      & 23.4         & 20.2          & 19.00     \\
      % Cardinal number fit
      \HCl     & 121          & 120           & 130.4     \\
    \end{tabular}
  \end{ruledtabular}
  \caption{Van der Waals constants $C_6^{\rm odd}$
           and $C_6^{\rm even}$ of the two-body dispersion interaction
           for the~\HF{} and \HCl{} chains. They are obtained by fitting
           the two-body energy increments of Fig.~\ref{fig:two-body-short}
           with either odd or even cardinal numbers~$n$, respectively,
           to Eq.~(\ref{eq:vdwdisp}).
           All data are given in~hartree~bohr$^6$.}
  \label{tab:van-der-Waals}
\end{table}

\subsection{Long-range correlations}
\label{sec:long}

\begin{figure}
  \includegraphics[width=\hsize,clip]{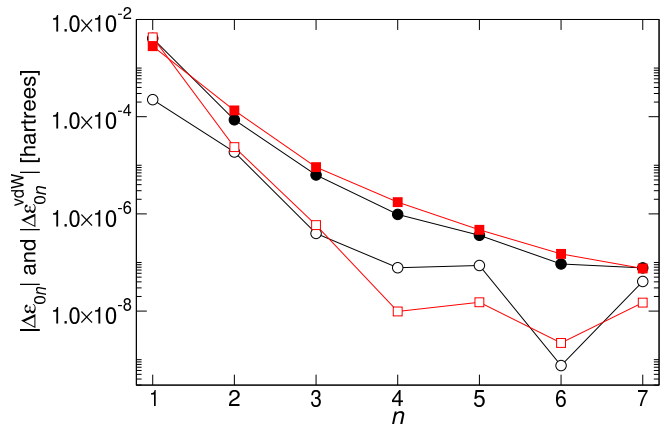}
  \caption{(Color online) Absolute values of two-body energy
           increments~$|\Delta \varepsilon_{0 \, n}|$ (closed symbols)
           and van der Waals-reduced two-body energy
           increments~$|\Delta \varepsilon_{0 \, n}^{\textrm{\scriptsize
           vdW}}|$ (open symbols)
           using Eq.~(\ref{eq:vdwdisp}) and the
           constants in Tab.~\ref{tab:van-der-Waals} for
           both the \HF{} chain (circles) and the \HCl{} chain (squares).}
  \label{fig:two-body-short}
\end{figure}

At a separation of two isolated monomers where the geometrical
overlap between the orbitals from distinct one-body orbital sets of the
monomers becomes negligible, only van der Waals dispersion
interactions remain.
This is also the case for the interaction between two
monomers in the infinite chains which is
correspondingly described by the two-body energy increments.
Their absolute values in~\HF{} and \HCl{}
for two monomers at a distance up to
seventh nearest neighbors are displayed in
Fig.~\ref{fig:two-body-short};
the curves drop off rapidly with distance.

The monomers in Fig.~\ref{fig:polygeom} that are labeled by
odd cardinal numbers are tilted with respect
to monomer~$0$. Likewise, monomers with even cardinal
numbers are arranged parallel to the monomer~$0$.
We will refer to the two types of two-body energy increments
that result from a parallel or tilted setting of monomers
as even or odd energy increments,
respectively.
The long-range interaction between two monomers
is approximated by the leading term of the two-body
van der Waals dispersion interaction~\cite{Klein:RG-76,Craig:MQ-84}
\begin{equation}
  \label{eq:vdwdisp}
  \varepsilon_{0 \, n}^{\textrm{\scriptsize vdW}} = \cases{
    \textstyle - \frac{C_6^{\rm odd }}{(\frac{n}{2} \, a)^6} \; , & $n$~odd       \cr
    \textstyle - \frac{C_6^{\rm even}}{(\frac{n}{2} \, a)^6} \; , & $n$~even$\;,$ \cr }
\end{equation}
where individual van der Waals constants~$C_6^{\rm odd}$ and
$C_6^{\rm even}$, respectively, are affixed for odd and even
energy increments.
Here $a$~represents the lattice constant of the chains.
%
% xmgrace 5.1.14 uses the Levenberg-Marquardt algorithm
% for non-linear curve fitting which is based on
% LMDIF from MINPACK, with some modifications.
%
% Literature (LMDIF):
% K. Levenberg (1944) A method for the solution of certain non-linear problems
% in least squares.  Quart. Appl. Math. 2:164-168.
%
% D.W. Marquardt (1963) An algorithm for least squares estimation of nonlinear
% parameters.  J. Soc. Industr. Appl. Math. 11:431-441  (also: SIAM J. Appl. Math.).
%
% Jorge J. More, The Levenberg-Marquardt Algorithm: Implementation
% and Theory. Numerical Analysis, G. A. Watson, editor.
% Lecture Notes in Mathematics 630, Springer-Verlag, 1977.
%
The van der Waals constants are obtained by a weighted
nonlinear curve fit~\cite{More:LM-78}
of the data in Fig.~\ref{fig:two-body-short} for either the odd
or the even energy increments.
The weights are chosen such that the fit reproduces the energy
increments with large~$n$ best, as for them the geometrical
orbital overlaps between the orbitals from two different
one-body orbital sets are negligible.
The van der Waals constants for the fits of the two sets
of translationally equivalent monomers are given in
Tab.~\ref{tab:van-der-Waals}.
Our theoretical data compare satisfactorily with the
experimental data for the dispersion interaction between
two isolated monomers also given in the table.

Having identified the van der Waals contribution to the
two-body energy increments in the chains, we can
subtract it
from the two-body energy increments to obtain
van der Waals-reduced energy increments
\begin{equation}
  \label{eq:vdw-reduced}
  \Delta \varepsilon^{\textrm{\scriptsize vdW}}_{0 n}
    = \Delta \varepsilon_{0 n} - \varepsilon_{0 n}^{
    \textrm{\scriptsize vdW}} \; .
\end{equation}
Their absolute value~$|\Delta \varepsilon_{0 n}^{
\textrm{\scriptsize vdW}}|$ is shown next to the absolute value of
the two-body energy increments~$|\Delta \varepsilon_{0 n}|$ in
Fig.~\ref{fig:two-body-short}.

Two regions can be identified in Fig.~\ref{fig:two-body-short}.
First, from the nearest to the third-nearest neighbors, there
is the local correlation zone, where, due to geometrical orbital
overlaps, short-range electron correlations are effective.
There, the decay of two-body energy increments
with the distance between the two monomers
is faster than the one that would result from a pure van
der Waals interaction, following Eq.~(\ref{eq:vdwdisp}).
We observe that the van der Waals contribution to~$|\Delta \varepsilon^{
\textrm{\scriptsize vdW}}_{0 1}|$ of~\HCl{} is appreciably
larger than in~\HF{} as the curves for~$|\Delta \varepsilon_{0 1}|$
and $|\Delta \varepsilon^{\textrm{\scriptsize vdW}}_{0 1}|$ are
much closer for~\HCl{} than for~\HF{}.
Second, there is the van der Waals zone reaching from the
third- up to the seventh-nearest neighbor.
Here a typical $r^{-6}$~decay is observed, leading to van
der Waals-reduced energy increments that are essentially zero.
A slight deviation of the two-body energy increments
from an $r^{-6}$~behavior is perceived beyond
fifth-nearest neighbors which likely can be
attributed to inaccuracies caused by the oligomer
approximation.
The translational relation of the Wannier orbitals is better
satisfied in~\HCl{} compared with~\HF{} [Sec.~\ref{sec:transfer}]
which leads to a lower absolute value of the fourth to seventh
van der Waals-reduced energy increments.
The low absolute value of~$\Delta \varepsilon^{\textrm{\scriptsize
vdW}}_{0 6}$ for both chains is an artifact of the fitting
process.
Figure~\ref{fig:two-body-short} also reveals that the two-body
energy increments
are of satisfactory accuracy even beyond the estimates
given in~Sec.~\ref{sec:transfer} of~$40 \U{\mu E_h}$ for~\oHF{10}
and $11 \U{\mu E_h}$ for~\oHCl{10}
due to a considerable error cancellation.

Given the van der Waals constants in Tab.~\ref{tab:van-der-Waals}, we
can use Eq.~(\ref{eq:vdwdisp}) to sum up the van der Waals
contribution of the two-body energy increments to the binding energy
to infinite distance~$\Sum_{n=3}^{\infty} \varepsilon_{0 n}^{
\textrm{\scriptsize vdW}}$ which yields
$-8 \U{\mu E_h}$~for~\HF{} and $-12 \U{\mu E_h}$~for~\HCl{}.
The bigger polarizability of chlorine atoms compared with
fluorine atoms leads to the larger contribution
in~\HCl{} when compared with~\HF{} as can be expected from the
analysis of the energy increments in Sec.~\ref{sec:increments}.
The contribution of two-body energy increments beyond
%
% HF: Amount of third-neighrest neighbor increment
% (-0.008 - (-0.0063)) / (-0.0063) = 0.2698
%
%
% HCl: Amount of third-neighrest neighbor increment
% (-0.012 - (-0.0092)) / (-0.0092) = 0.3044
%
third-nearest neighbors is in both cases~$\approx 30 \%$
of~$\Delta \varepsilon_{0 \, 3}$.

Generally, once a good estimate of the van der Waals constants of a crystal is
available, the decomposition of the energy increments
into a short- and a long-range van der Waals
contribution can be made. This allows one to
focus on the short-range part.
Thereby, the required number of energy increments
to be calculated to reach a certain accuracy of the correlation
energy is reduced considerably.
Therefore, the computational demand of its determination is
also significantly decreased.

\section{Binding energy}
\label{sec:bind}

Having understood the impact of electron correlations
in~\HF{} and \HCl{} chains, we are now in a position to examine
the total binding energy per monomer~$\Delta E$ of the chains;
it is given by
\begin{equation}
  \label{eq:RHF_bind}
  \begin{array}{rcl}
    \Delta E &=& \Delta E_{\rm SCF} + \Delta E_{\rm corr} \\
             &=& \frac{1}{2} \, (\Cal E^{\rm SCF}_{\rm chain}
                 + \Cal E^{\rm corr}_{\rm chain})
                 - \Cal E^{\rm SCF}_{\rm monomer}
                 - \Cal E^{\rm corr}_{\rm monomer} \; .
  \end{array}
\end{equation}
It consists of the Hartree-Fock and the correlation
contributions~$\Delta E_{\rm SCF}$ and $\Delta
E_{\rm corr}$, respectively, where $\Cal E^{\rm
SCF}_{\rm chain}$ and $\Cal E^{\rm corr}_{\rm chain}$
denote the Hartree-Fock and correlation energies per
unit cell of the chains.
Furthermore, $\Cal E^{\rm SCF}_{\rm monomer}$
and $\Cal E^{\rm corr}_{\rm monomer}$ are the corresponding
energies for the monomers.
The Hartree-Fock energies are readily available from molecular and
periodic calculations~\cite{Pisani:HF-88,Pisani:QM-96,crystal03}
but electron correlations are more involved and were discussed
in the previous Sec.~\ref{sec:elcorr}.

In Tab.~\ref{tab:bindtotal}, we communicate basis-set-extrapolated
Hartree-Fock binding energies~$\Delta E_{\rm SCF}(\infty)$ for the
chains taken from Refs.~\onlinecite{Buth:BS-04,Buth:MC-05}.
At the Hartree-Fock equilibrium geometry of~\HF{},
Bayer and Karpfen give~$-10.360 \mHartree$ for basis set~3 in
Ref.~\onlinecite{Karpfen:AI-82} while Hirata and Iwata
find~$-10.855 \mHartree$ for the 6-311++G(d,p) basis
set.~\cite{Hirata:AI-98}
Both numbers are in good agreement with our result which
is~$-10.202 \mHartree$.
Yet Berski and Latajka report~$-9.696 \mHartree$ for~\HF{},
when using the 6-311G(d,p) basis set,~\cite{Berski:PHF-97} and
$-2.073 \mHartree$ for~\HCl{} harnessing the DZ(d,p)
basis set.~\cite{Berski:PHF-98}
The latter number in particular deviates considerably from our
result, $-0.586 \mHartree$,
which is most likely due to the fact that
they did not remove the BSSE.~\cite{Berski:PHF-97,Berski:PHF-98}

To determine the correlation energy per monomer of the
infinite chains~$\Cal E_{\rm corr}$, we inspect
the energy increments in Tab.~\ref{tab:increments}.
They converge rapidly both with increasing distance among the
monomers involved and with increasing order of the many-body
expansion, \ie, with the maximum number of monomers correlated
in a specific energy increment.
In order to obtain an accuracy in~$\Cal E_{\rm corr}$
of $\approx 50 \U{\mu E_h}$ for~\HF{} and
of $\approx 11 \U{\mu E_h}$ for~\HCl{},
we see from Tab.~\ref{tab:increments} that it is
enough to include the following energy increments:
\begin{equation}
  \label{eq:corrpoly}
  \Cal E_{\rm corr} \approx \Delta \varepsilon_0 + \Delta \varepsilon_{01}
  + \Delta \varepsilon_{02} \; .
\end{equation}
Hence, it is sufficient to correlate
the electrons in the Wannier orbitals of only two
monomers at a time for \HF{} and \HCl{}~chains
in which more distant two-body energy increments contribute
only $-8 \U{\mu E_h}$~for~\HF{} and $-12 \U{\mu E_h}$~for~\HCl{}
as shown in Sec.~\ref{sec:long}.
In three-dimensional crystals, three-body terms
become more important according to the analyses in
Refs.~\onlinecite{Axilrod:IV-43,Fink:AI-95,Rosciszewski:AI-00}
and the much larger number of three-body energy increments
compared with the number of two-body energy increments.

We report~$\Delta E$ based on the approximation
of Eq.~(\ref{eq:corrpoly}) as obtained by basis-set
extrapolation~\cite{Buth:BS-04,Buth:MC-05} in Tab.~\ref{tab:bindtotal}.
Using density functional theory, Hirata and Iwata~\cite{Hirata:AI-98}
find at optimized geometries for the total binding energy
of~\HF{} $-13.521 \mHartree$ (BLYP) and $-13.864 \mHartree$ (B3LYP)
[utilizing the 6-311++G(d,p) basis set] which are larger in magnitude
by~$14\%$ and $17\%$, respectively, than our result
for the experimental geometry.
The observed differences can be partially ascribed
to the different geometries used;
our experimental geometry leads to a binding energy that is
somewhat smaller in magnitude than the binding energy in
the optimized geometry.
The 6-311++G(d,p) basis set of Iwata and Hirata~\cite{Hirata:AI-98} also
differs substantially from our choices
in Refs.~\onlinecite{Buth:BS-04,Buth:MC-05}.
Above all, we do not consider the impact of triple excitations in our study;
they increase the absolute value of the short-range correlation
contributions and, consequently, also the absolute value of
the binding energies.

\section{Conclusion}
\label{sec:conclusion}

\begin{table}
%
% HF Hartree-Fock binding energy per monomer: -11.82525 mEh
% Mean of all least squares fitted Hartree-Fock binding energies:
% ((-10.199) + (-10.194) + (-10.194) + (-10.222)) / 4 = -10.20225 mEh = -0.27761735 eV
%
% Mean of the best extrapolated CCSD binding energies:
% ((-1.655) + (-1.606) + (-1.646) + (-1.585)) / 4 = -1.623 mEh = -0.044164 eV
%
%
%
% HCl Hartree-Fock binding energy per monomer: -3.34975 mEh
% Mean of all least squares fitted Hartree-Fock binding energies:
% ((-0.596) + (-0.586) + (-0.584) + (-0.577)) / 4 = -0.58575 mEh = -0.0159391 eV
%
% Mean of the best extrapolated CCSD binding energies:
% ((-2.773) + (-2.747) + (-2.774) + (-2.762)) / 4 = -2.764 mEh = -0.075212 eV
%
%
  \centering
  \begin{ruledtabular}
    \begin{tabular}{rdd}
                   & $\HF{}$ & $\HCl{}$ \\
      \hline
      $\Delta E_{\rm SCF}(\infty)$  & -10.202 & -0.586 \\
      $\Delta E_{\rm corr}(\infty)$ &  -1.623 & -2.764 \\
      $\Delta E(\infty)$            & -11.826 & -3.350 \\
    \end{tabular}
  \end{ruledtabular}
  \caption{Basis-set-extrapolated total binding energies per
           monomer~$\Delta E(\infty)$ of~\HF{} and \HCl{}~chains and their
           decomposition into basis-set-extrapolated
           Hartree-Fock~$\Delta E_{\rm SCF}(\infty)$ and
           electron correlation~$\Delta E_{\rm corr}(\infty)$
           contributions.
           All data are taken from Refs.~\onlinecite{Buth:BS-04,Buth:MC-05}
           and are given in millihartrees.}
  \label{tab:bindtotal}
\end{table}

We study hydrogen bonding in infinite \HF{} and \HCl{} chains.
Whereas in~\HF{} the Hartree-Fock contribution dominates the total
binding energy
%
% -10.202 / -11.826 = 86.27%
%
by~$86\%$ due to the electrostatic contribution of the rather ionic
HF~monomers, in the much more weakly bound~\HCl{}, the Hartree-Fock
calculation yields
%
% -0.586 /  -3.350 = 17.49%
%
only~$18\%$ of the binding energy [Tab.~\ref{tab:bindtotal}].
The transitional character of hydrogen-bonded crystals between the
ionic and van der Waals regimes is reflected in these numbers.
The very weak hydrogen bonds in the \HCl{} chains bear a
close resemblance to purely van der Waals-bonded systems,
like rare-gas solids, where
bonding is entirely caused by electron
correlations.~\cite{Fink:AI-95,Rosciszewski:AI-99,%
Rosciszewski:AI-00}
In fact only the inclusion of electron correlations
puts the binding energy per monomer of the \HCl{} chain
into the range conventionally ascribed to hydrogen
bonding,~\cite{Pauling:NC-93}
%
% Pauling gives for hydrogen bonding the range 2 to 10 kJ / mol
% per bond.
%
%  2 kJ / mol =  2 * 4.184 / 96.485 / 27.2113845 * 1000 mE_h
%            =  3.187 mE_h
%
% 10 kJ / mol = 10 * 4.184 / 96.485 / 27.2113845 * 1000 mE_h
%            = 15.936 mE_h
%
namely~\hbox{$3$--$16 \mHartree$}.

The incremental scheme, which provides a decomposition of the
contribution of electron correlations to the binding energy of the
chains in terms of nonadditive many-body energy increments,
is shown to converge rapidly with respect to the number
of monomers, \ie, bodies, involved in the energy increments
and the distance among them, thus providing a
good tool to study hydrogen bonding in crystals.
In contrast to three-dimensional crystals, the dominant
contribution of electron correlations to the binding energies of
the infinite chains is, in the present investigation,
already given by pair interactions
between the valence electrons of a monomer and its nearest
and next-nearest neighbors.
The $K$-body energy increments represent a general framework
to describe van der Waals dispersion
interaction among the electrons from $K$~one-body orbital sets.
The decay of the two-body energy increments,
\ie, the electron correlations between two monomers in the
chains, with the intermonomer distance is investigated and
van der Waals constants are determined,
which affords a partition of the energy increments
into a short-range, van der Waals-reduced part
and a long-range part that can be summed analytically
to infinite distances.

This partitioning affords a methodological advancement;
if the van der Waals constants of a crystal
are available from other sources, then the number of energy increments
in the incremental expansion can be reduced by considering only
van der Waals-reduced energy increments;
they typically need to comprise only the nearest- and
next-nearest-neighbor energy increments.
The small number of van der Waals-reduced energy increments
permits one to significantly decrease the number of energy
increments which have actually to be calculated
to reach a certain accuracy of the correlation energy
compared with the conventional incremental series.
Furthermore, the long-range part turns out to be
well described using small (double-$\zeta$) basis
sets---in contrast to the short-range part where basis-set
extrapolation is required for an accurate
description~\cite{Buth:BS-04,Buth:MC-05}---thus facilitating
further computational savings.

\begin{acknowledgments}
We are highly indebted to Uwe Birkenheuer, Krzysztof Ro\'sciszewski,
Hermann Stoll, and Peter Fulde for helpful discussions and
a critical reading of the manuscript.
\end{acknowledgments}

\end{document}